\documentstyle[twoside,fleqn,espcrc2,epsfig]{article}

\newcommand{\beq}{\begin{equation}}
\newcommand{\eeq}{\end{equation}}
\newcommand{\beqn}{\begin{eqnarray}}
\newcommand{\eeqn}{\end{eqnarray}}

\def\NP{{\it Nucl. Phys.\,}}
\def\PL{{\it Phys. Lett.\,}}

\def\PR{{\it Phys. Rev.\,}}

\title{
\vspace{-3.6cm}
\begin{flushright}
{\normalsize
ITEP-TH-43/97\\
\vspace{-.2cm}
September 97}
\end{flushright}
\vspace{1.5cm}
Abelian monopoles in lattice gluodynamics as physical objects
\thanks{Talk given by M.I.~Polikarpov at
the International Symposium on Lattice Field Theory, 22-26 July
1997, Edinburgh, Scotland}}

\author{B.L.G.~Bakker\address{Department of Physics and Astronomy,
Vrije Universiteit,\\
De Boelelaan 1081, NL-1081 HV Amsterdam, The Netherlands},
M.N.~Chernodub\address{ITEP, B.Cheremushkinskaya 25, Moscow,
117259, Russia} and
M.I.~Polikarpov$^b$}

\begin{document}

\begin{abstract}
By numerical calculations we show that the abelian monopole
currents are locally correlated with the density of $SU(2)$ lattice
action. This fact is established for the maximal abelian projection.  Thus,
in the maximal abelian projection the monopoles are  physical
objects, they carry $SU(2)$ action. Calculations on the
asymmetric lattice show that the correlation between monopole
currents and the density of $SU(2)$ lattice action also  exists
 in the deconfinement phase of gluodynamics.
\end{abstract}

\maketitle

\section{INTRODUCTION}

The monopoles in the maximal abelian projection (MaA projection) of
$SU(2)$ lattice gluodynamics \cite{KrScWi87} seem to be responsible
for the formation of the flux tube between the test quark-antiquark
pair. The $SU(2)$ string tension is well described  by the
contribution of the abelian monopole currents
\cite{ShSu94}  which satisfy the London equation
for a superconductor \cite{SiBrHa93}. The study of monopole creation
operators shows that the abelian monopoles are condensed
\cite{DiGi95} in the confinement phase of
gluodynamics.

\mbox{}  On the other hand,  the
abelian monopoles arise in the
continuum theory \cite{tHo81} from the singular gauge transformation
and it is not clear whether these monopoles are ``real'' objects. A
physical object is something which carries action and in the present
publication we study the question if there are any correlations
between abelian monopole currents and $SU(2)$ action. In
 \cite{ShSu95} it was found that the total action of $SU(2)$
fields is correlated with the total length of the monopole currents,
so there exists a global correlation. Below we discuss the local
correlations between the action density and the monopole currents.

\section{CORRELATORS OF MONOPOLE CURRENTS AND DENSITY OF $SU(2)$
ACTION}

The simplest quantity which reflects the correlation of the local
action density and the monopole current is the relative excess of
$SU(2)$ action density in the region near the monopole current. It
can be defined as follows. Consider the average action  $S_m$  on the
plaquettes   closest to the monopole current $j_\mu(x)$. Then
the relative excess of the action is

\beq
\eta = \frac{S_m - S}{S}\;, \label{eta}
\eeq
where $S$ is the standard expectation value of the lattice action, $S
=\left <\left( 1 - \frac 12 Tr \, U_{P}\right)\right>$, $S_m$ is
defined as follows:

\beq
S_m= \left<\frac 16
\sum_{P \in \partial C_\nu (x)}
\left( 1 - \frac 12 Tr \, U_P \right)\right>\, , \label{Smlat}
\eeq
were the average is implied over all cubes $C_\nu (x)$
dual to the magnetic monopole currents $j_\nu(x)$, the summation is
over the plaquettes $P$ which are the faces of the cube $C_\nu (x)$;
$U_P$ is the plaquette matrix. For the static monopole
 {  we have} $j_0 (x) \neq
0,\, j_i (x) =0, \, i=1,2,3$;  and only the  magnetic part of $SU(2)$
action density contributes to $S_m$.

\section{NUMERICAL RESULTS}

We calculate the quantity $\eta$ on the symmetric lattice
($10^4$) and on the lattice $12^3 \cdot 4$ which corresponds to finite
temperature. In both cases, it occurs that  in the MaA projection we have
  $\eta \neq 0$ for all
considered values of $\beta$. We
also study  the abelian projection  which corresponds to the
diagonalization of the plaquette matrices in the 12 plane ($F_{12}$
gauge) and to the diagonalization of the Polyakov line (Polyakov
gauge).

\begin{center}
\begin{figure}[htb]
\epsfig{file=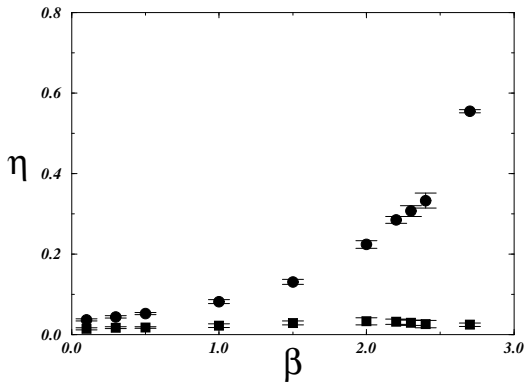,height=5.0cm}
\vspace{-.3cm}
\caption{The relative excess of the magnetic action density near the
monopole current, $\eta$, for the $10^4$ lattice. Circles correspond
to MaA projection, squares correspond to Polyakov gauge.}
\vspace{-.8cm}
\end{figure}
\end{center}

In Fig.1,  we show  the dependence of the quantity $\eta$ on $\beta$ for
the lattice of size $10^4$ for the MaA projection and for the Polyakov
gauge. It turns  out that the data for the $F_{12}$ projection
coincide within  statistical errors with the data for the Polyakov
gauge and we do not show  them. In Fig.~2,  we plot the same data, this time,
for the asymmetric lattice $12^3\cdot 4$. It is seen that the quantity
$\eta$ is much smaller for the Polyakov gauge than for the MaA
projection; the deconfinement phase transition at $\beta
\approx 2.3$ does not have much  influence on the behavior of $\eta$.
Thus,  the
monopole currents in the MaA projection are surrounded by plaquettes
which carry the values of $SU(2)$ action  larger than the value of the
average plaquette action.

To obtain these results we consider 24 statistically independent
configurations of $SU(2)$ gauge fields for $\beta \le 2.0$, 48
configurations for $2.25 \le \beta \le 2.35$,  and 120 configurations
for $\beta \ge 2.4$. To fix the MaA projection we  have used the
overrelaxation algorithm \cite{MaOg90}. The number of the gauge
fixing iterations is determined by the criterion  given in
\cite{Pou97}: the iterations are stopped when the matrix of
the gauge transformation  $\Omega (x)$  becomes close to the unit
matrix: ${\rm max}_x \left\{ 1- {\frac 12} Tr \,\Omega (x) \right\} \le
10^{-5}$.  It has been  checked that more accurate gauge fixing does not
change our results.

\begin{center}
\begin{figure}[htb]
\epsfig{file=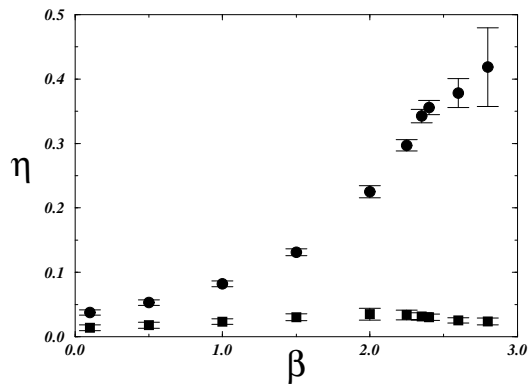,height=5.0cm}
\vspace{-.3cm}
\caption{The same as in Fig.~1, but for the asymmetric lattice
$12^3 \cdot 4$.}
\vspace{-.8cm}
\end{figure}
\end{center}

Thus we have shown that in the MaA projection the {\it abelian}
monopole currents are surrounded by regions with a high {\it
nonabelian} action. This fact means that the monopoles in the MaA
projection are physical objects. It does not mean that they have to
be real objects in the Minkovsky space. What we have  found is that these
currents carry $SU(2)$ action in the Euclidean space. It is
important to understand what is the general class of configurations
of $SU(2)$ fields which generate the monopole currents. Some
specific examples are known. These  are instantons
\cite{GuCh95} and the  BPS--monopoles
(periodic instantons) \cite{SmSi91}. This question can be
formulated in another way: are there any continuum physical objects
which correspond to the  abelian monopoles obtained in the MaA projection?

\section*{ACKNOWLEDGMENTS}
%\vspace*{0.8cm} %\noindent
%{\normalsize\bf Acknowledgments}\par\vspace*{0.4cm}

The authors are grateful to   Prof.  T.~Suzuki for useful
discussions. M.I.P. and M.N.Ch. feel much obliged for the kind
reception given to them by the staff of the Department of Physics of
the Kanazawa University, and by the members of the Department of
Physics and Astronomy of the Free University at Amsterdam. This work
was supported by the JSPS Program on Japan -- FSU scientists
collaboration, by the grants INTAS-94-0840, INTAS-94-2851,
INTAS-RFBR-95-0681 and RFBR-96-02-17230a.

\end{document}